# A reaction-diffusion model for the progression of Parkinson's disease[*]

Míriam R. García[1], Mathieu Cloutier[2], Peter Wellstead[3][†]

[1] Bioprocess Engineering Group. IIM-CSIC, Vigo Spain.
To whom correspondence should be addressed (miriam@iim.csic.es)
[2] Ecole Polytechnique de Montreal, Montreal, QC, Canada
[3] NUIM, Maynooth, Ireland

**The temporal and spatial development of Parkinson's disease has been characterised as the progressive formation of α-synuclein aggregations through susceptible neuronal pathways. This article describes a new model for this progression mechanism in which Parkinsonian damage moves over time through the nervous system by the combined effect of the reaction kinetics of pathogenesis and molecular diffusion. In the reaction-diffusion model, the change from a healthy state to the disease state advances through the nervous system as a wave front of Parkinsonian damage, marking its path by accumulations of damaged α-synuclein and neurotoxic levels of oxidative species. Progression according to this model follows the most vulnerable routes through the nervous system as described by Braak's staging theory and predicts that damage will advance at differing speeds depending upon the level and number of risk factors, in a manner that gives new insights into the variations with which Parkinson's disease develops.**

## I. INTRODUCTION

The pathology of Parkinson's disease is characterised by the progressive appearance in the nervous system of aggregations (consisting principally of damaged α-synuclein) that follow '*an ascending course with little inter-individual variations*' [Braak, 2003a]. The same authors have suggested that this spatially progressive sequence could be achieved [Braak, 2003b] by an invading pathogen which enters the enteric nervous system (ENS) and proceeds by backward projection to the brain stem and hence the brain itself. The Braak progression theory concerns the long term development of Parkinson's disease, and is therefore highly important in building an overall understanding of Parkinson's from beginning to end.

The idea that an invading pathogen can itself propagate through the nervous systems has been thrown into question by the finding [Pan-Montojo, 2010] that Parkinsonian damage an animal model caused by intragastrically administered toxin is not accompanied by traces of the toxin. This paper describes an explanation of progression that overcomes this problem. Specifically, we propose that, while an invading pathogen may initiate Parkinsonian damage locally, it is the endogenous pathogenesis mechanism of the disease, rather than the pathogen, that causes subsequent transmission and progression.

In particular, we describe a model for the spread of Parkinsonian damage by a combination of (i) the reaction kinetics of pathogenesis which cause the local growth of α-synuclein misfolds (αSYNmis) and the rise of reactive oxidative species (ROS) to neuro-toxic levels, plus (ii) molecular diffusion of these pathogenic products. We envisage the process as operating as follows: an initial trigger mechanism (e.g. an external pathogen or other stress factor) overthrows the homeostatic neurochemical balance within a localised compartment somewhere in the nervous system. This causes local pathogenesis whereby levels of α-synuclein increase and reactive oxidative species (ROS) rise to neurotoxic levels that create the disease state.

In this model it is the neurochemical reaction dynamics of pathogenesis, combined with molecular diffusion, which communicates the elevated ROS and protein damage to neighbouring compartments. This causes the effected compartments themselves undergo pathogenesis, such that Parkinsonian damage spreads sequentially from compartment to compartment. The rate of progression of the disease state depends upon the kinetics of the pathogenic reaction and this, in turn, is determined by the severity of the risk factors and the vulnerability of individual neurons.

## II. A MODEL FOR THE PATHOGENESIS OF PARKINSON'S DISEASE

### A. *Pathogenesis as a critical transition within a neuronal compartment.*

Despite being a well-known phenomenon in biology [Murray, 2002, Volpert, 2009], the spread of chemical species and concentrations by reaction-diffusion (R-D) has not previously been considered as a progression mechanism for Parkinson's disease. This is probably because the neurochemical kinetics of Parkinson's pathogenesis has not previously been understood (see, for example, [Lotharius, 2002, Hattori, N, 2004, Shin, 2009, Schapira, 2011]). However, a recently developed mathematical model [Cloutier, 2012a] of α-synuclein and oxidative metabolism simulates the pathogenesis of Parkinson's within a neuronal compartment, and gives an analytical picture of the pathogenic process [Cloutier, 2012b]. Specifically, when placed under sufficient stress from Parkinsonian risk factors, the model exhibits a sudden (and apparently spontaneous) transition from healthy neurochemical conditions to a disease state that creates Parkinsonian damage within the neuronal compartment. The abrupt transition from healthy to a disease state occurs when risk factors acting on the neuronal compartment exceed a critical level. Beyond this critical level, a bifurcation occurs that cause the concentrations of misfolded α-synuclein (αSYNmis) and reactive oxygen species (ROS) to undergo what is known in natural sciences as a critical transition [Scheffer, 2009], as they increase to high levels. In most cases, but not necessarily, the switch to neuro toxic levels is permanent, persisting even after the stress initiating it has been removed [Cloutier, 2012b]. (See Supplementary Materials SM2 for a simplified model of the pathogenic mechanism). This model of Parkinson's pathogenesis arose from mathematical modelling of brain energy metabolism [Cloutier, 2012c], and how it interacts with the neurochemical dynamics of reactive species and α-synuclein metabolism. However, although it is a mathematical modelling exercise, the models were calibrated using measured changes in neurochemicals [Lowry, 2004], and the pathogenesis finding has been investigated in animal models. Specifically, Paraquat was administered periodically to freely moving rats in a manner used in simulations, and continuous long term measurement of cerebral nitric oxide (NO) reproduced the predicted transition to a Parkinsonian state [Finnerty 2013].

---

[*] In memorial to Prof. Peter Wellstead. This was his last piece of work.
[†] Deceased 24th June 2016.

*B. External stress as an initiator of the critical transition to Parkinson's disease.*

To illustrate how an external pathogen might initiate Parkinson's disease, we start with the example used by Braak [Braak, 2003b], but in a much simplified form. Specifically, we consider that the start point is in the ENS and this is connected to the brain stem by a single neuronal pathway in which all parts of the neuronal tissue have uniform properties. Figure 1 illustrates this situation where the invading pathogen is shown as an external toxic stress and the approximate distance from ENS to the brain stem is 1 metre.

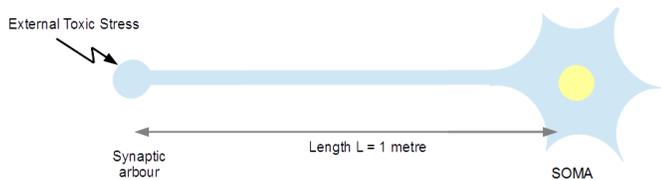

Figure 1: A hypothetical continuous neuronal pathway from ENS to brain stem

Figure 2 shows what happens in the model of pathogenesis when an invading pathogen (in this case a transient loading of external toxin) enters the compartment of the simulated neuron at its closest point to the pathogens point of entry. Figure 2 shows the permanent change in the steady state level of reactive oxygen species and α-synuclein misfolds that is caused by the toxic stress. Below a critical level of external stress (duration and magnitude), the compartment is not switched permanently and levels of ROS and αSYNmis recover to their previous levels (see simulations in [Cloutier, 2012a, b] for details). This is compatible with practical observations on the impact of low or infrequent exposures to certain chemical used in agricultural and industrial applications.

The external stress in Figure 2 exceeds the critical point that causes a bifurcation [Kuznetsov, 2004] to occur in the nonlinear kinetics of the ROS/ $αSYN_{mis}$ interaction, resulting in the steady state levels of ROS and $αSYN_{mis}$ to permanently increase to levels where the development of Parkinson's disease is initiated by α-synuclein accumulation and neurotoxic oxidative damage. The external toxic stress used in this simulation is transient, lasting only 40 days (from day 10 to day 50). A transient external stress is simulated to emphasize the point that, despite the removal of the initiating stress, the change to a Parkinsonian state in the neuronal compartment remains. That is to say, the external stress/invading pathogen effectively operates a one-way switch from healthy homeostasis to a stable disease state in the neuronal compartment.

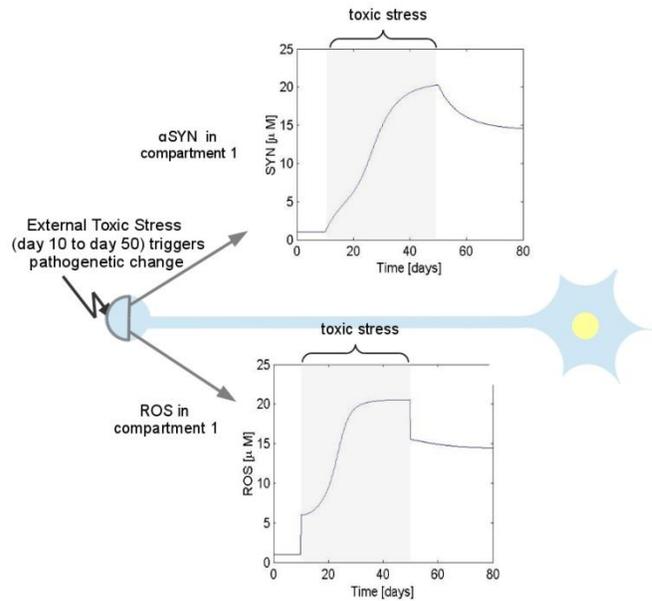

Figure 2: Concentrations of ROS and αSYNmis in compartment 1 switch from low healthy levels to high levels triggered by exposure to external toxic stress from day 10 to day 50.

*C. Transmission of the disease state to adjacent neuronal compartments*

Figure 3 illustrates the onward transmission of high levels of ROS and αSYNmis in terms of how the pathogenic switch to high levels of ROS and αSYNmis in compartment 1, then causes a similar transition to be triggered in the immediately adjacent area (compartment 2). Parkinsonian damages thus progresses from the first to the second compartment without the initiating pathogen. This process is illustrated in Figure 3 where the disease state from the compartment 1 (ROS 1/ αSYNmis 1), triggers a switch in compartment 2 to a Parkinsonian state with elevated concentrations (ROS 2/ αSYNmis 2).

Figure 3 illustrates the point that the initial trigger mechanism for Parkinson's need only be the initial agent for pathogenic change and does not progress further. The separation between the initiating agent at the disease start point and the subsequent spread by endogenous pathogenic transition rationalises the results [Pan-Montojo, 2010].

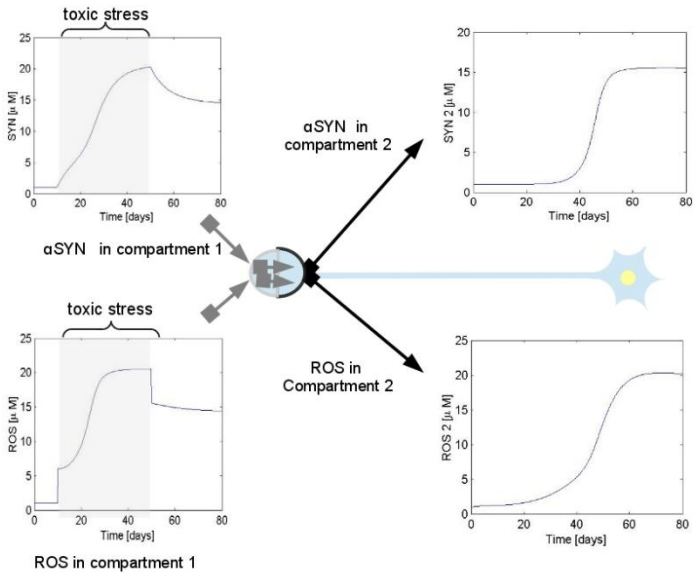

Figure 3: The raised levels of concentrations of ROS and αSYNmis in compartment 1 trigger a corresponding switch in ROS and αSYNmis in compartment 2.

### III. A REACTION-DIFFUSION MODEL OF PARKINSON'S PROGRESSION

The explanation in the previous paragraphs can be extended to model a continuous pathway through the nervous system as a set of contiguous neuronal compartments in which the pathogenic transition is communicated from one compartment to the next, like a line of falling dominos. However, this 'passing on' of the disease state from one compartment to another is more elegantly (and correctly) represented mathematical by considering the continuous neural pathway to consist of a sequence of neuronal compartments which become thinner and thinner. In the limit, as the compartments become infinitesimally thin, the model becomes a partial differential equation [Murray, 2002] that combines molecular diffusion of ROS and αSYNmis with the reaction equations for pathogenesis (see Supplementary Material SM2 for details of the reaction-diffusion equations for Parkinson's progression).

#### A. Simulating reaction-diffusion progression from the ENS to the brain stem.

To illustrate how reaction-diffusion would work as a progression mechanism, we simulate the progression of Parkinsonian damage described in [Braak, 2003b] from the ENS to the brain stem by computing the Parkinsons's reaction-diffusion equation along the hypothetical 1 metre neuron of Figure 1. Since the intention is to the principle of how Parkinson's can progress by reaction-diffusion, we assume that the neuron has uniform properties along its entire length.

The computation begins with initial conditions in which ROS and αSYN concentrations are at levels compatible with a healthy neuronal state along the full one metre length of the hypothetical neuron (Figure 1). An initial external toxic stress is applied to the start point (the synaptic arbour in Figure 1) and the partial differential equations which govern variations in concentrations of ROS and αSYNmis are calculated (See SM2) at each point along the length over a 20 year time period.

Consider first a neuron with no external risks from excessive aging, head trauma, genetic disorder or toxic damage (apart from the initial toxin exposure at the start point).

Figure 4 shows the development in the concentrations of ROS and αSYNmis over time and distance from the start point for this situation. As can be seen, the switch to a disease state propagates from the start point at time zero in the form of a wave front. It finally reaches the end point (the soma) approximately 7 years after toxic stress initiated the disease state at the start. This propagation time is of the same order of magnitude as might be expected for the progression of Parkinsonian damage in reality. For the purposes of comparison with the alternative of diffusion alone, the progression over the same length by diffusion of ROS and αSYNmis, would be much slower. Using published data for diffusion rates it would take approximately 300 years for Parkinson's to progress the one metre from ENS to brain stem. (See Supplementary Material SM2). Thus reaction-diffusion gives a plausible progression time for Parkinson's initial step from ENS to brain stem – now we consider how changes in known Parkinsonian risk factors might change the speed of progression of the disease

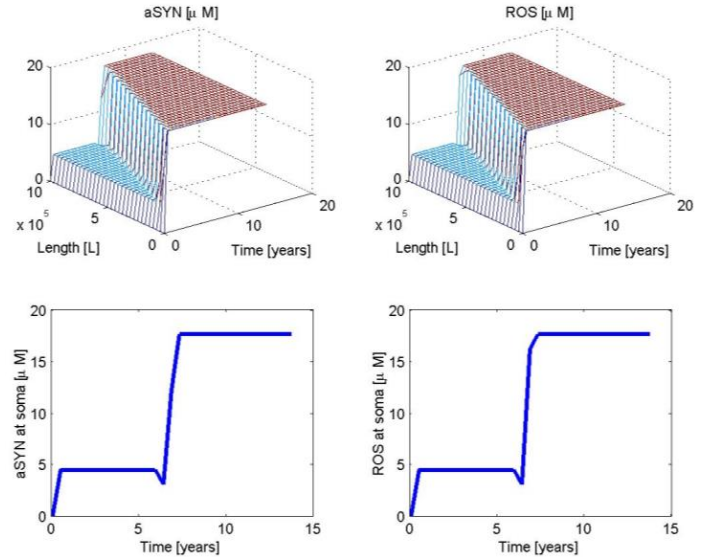

Figure 4: Profile over time and length of a hypothetical 1 metre neuron with progression by reaction-diffusion

### IV. THE IMPACT OF RISK FACTORS ON THE PROGRESSION OF PARKINSON'S

In the mathematical model that simulates pathogenesis, the speed and magnitude of the transition from healthy to disease state is strongly influenced by the size and number of Parkinsonian risk factors [Cloutier, 2012c, Wellstead, 2012]. In this section we demonstrate that this influence transfer to the reaction – diffusion model of progression, such that the speed with which Parkinson's progresses is also a direct function of Parkinsonian risk factors. We begin with the influence of aging on progression speed.

## A. Ageing

Ageing is the main risk factor for developing Parkinson's disease, and has a strong influence on the model of the pathogenic mechanism for local transition to the disease state. It is therefore of great interest to see if this influence extends to rate at which of Parkinsonian damage progresses through the nervous system. This can be illustrated by adjusting the aging rate in the reaction-diffusion equations – the level of aging is modelled by the strength of the brain energy metabolism.

Figure 5 (i) shows the progression of Parkinsonian damage for a higher aging rate than normal (weaker brain energy metabolism). In this case Parkinsonian damage progresses the one metre along the model neuron in approximately 3.5 years, significantly faster than in the normal aging case of Figure 4.

The converse is also true and low age effect (stronger brain energy metabolism) slows progression. Figure 5(ii) illustrates how a reduced the aging level also reduces the speed of progression: in the simulation shown Parkinsonian damage has not reached the soma 15 years after the initial toxic stress at the ENS.

Our models do not precisely classify the aging rate. Nonetheless, the trends in the progression (slower for low aging rate – faster for high aging rate) are consistent with the variable nature of progression observed in practice. It also raises point that the ENS may receive frequent damaging toxic insults which, because of a strong brain energy metabolism never progress sufficiently fast to the brain.

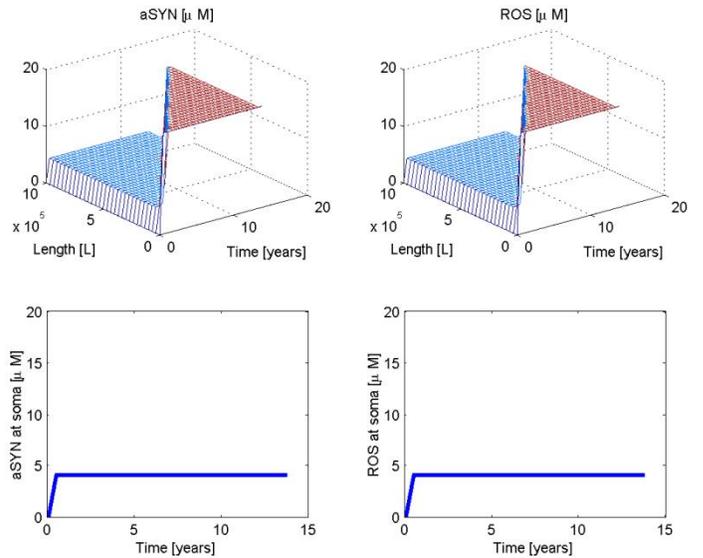

Figure 5 (ii) : Profile over time and length of a hypothetical 1 metre neuron with progression by reaction-diffusion with low aging (strong brain energy metabolism)

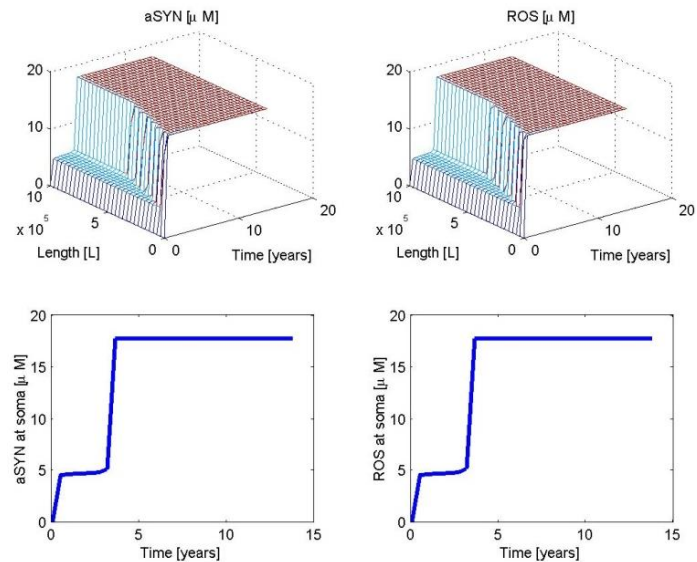

Figure 5 (i) : Profile over time and length of a hypothetical 1 metre neuron with progression by reaction-diffusion with high aging (weak brain energy metabolism)

## B. Genetic predisposition

Parkinson's disease is only a truly genetic condition in a minority of cases. Nonetheless, mutations in certain genes create a predisposition to Parkinson's disease. This is reflected in the model of pathogenesis (SM2) in which increased genetic irregularity lowers the switching threshold to the Parkinsonian state and increases the toxicity of the disease state, [Cloutier, 2012c]. Thus genetic disorders are predicted to make Parkinson's more likely and more severe.

The corresponding impact of genetic disorder on the speed of progression is simulated in Figure 6 (i) which shows that for a modest level of genetic mutation, the time to progress from synaptic terminal to the soma is reduced from 7 years, to approximately 3.5 years. Figure 6 (ii) shows the corresponding plot for a high level of genetic irregularity (corresponding to familial Parkinson's) in which the progression time over the 1 metre length is reduced to less than one year.

The large variability in progression times with different risk factors (in this case with age and genetic predisposition, but in practice with other risk factors as well) is in line with, and gives potential reasons for, the great variabilities observed in the severity and progression of Parkinson's in practice.

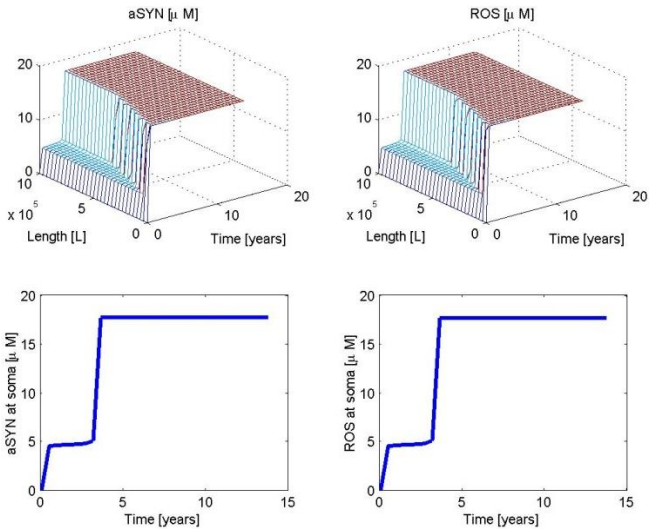

Figure 6 (i) : Profile over time and length of a hypothetical 1 metre neuron with progression by reaction-diffusion with moderate genetic predisposition

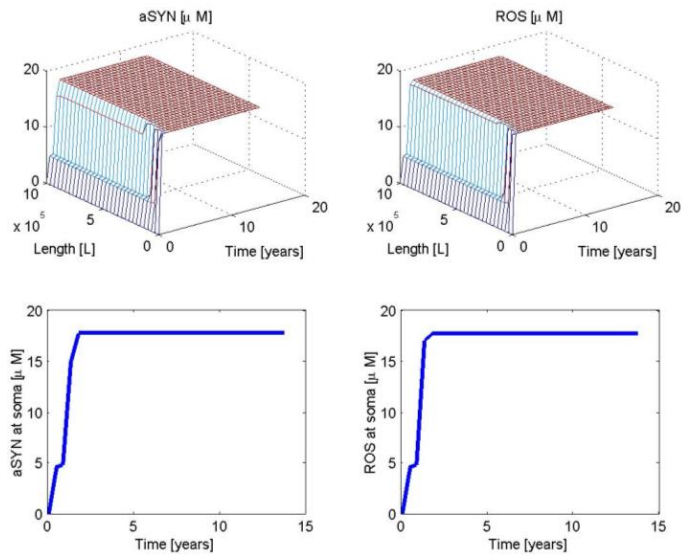

Figure 6 (ii) : Profile over time and length of a hypothetical 1 metre neuron with progression by reaction-diffusion with strong genetic predisposition

### C. Sensitivity to diffusion rates: α-synuclein

The protein α-synuclein is an unstructured protein which can diffuse at various rates depending on its conformation [Ducas 2014]. Likewise, uncertainty of the role of active transport of α-synuclein in misfolded form, lead us to consider the sensitivity of progression in the R-D model to variations in diffusion rate. In particular, we computed the R-D model for a range of α-synuclein diffusion rates. The results show that the progression time by reaction-diffusion is indeed sensitive to the diffusion rate of α-synuclein. The nominal diffusion rate used in the R-D computations (see SM2) is $D_a = 72000$ m$^2$h$^{-1}$ and gave a progression time of approximately 7 years. Increasing the diffusion rate by a factor of four (the range given in [Ducas 2014]) gives a faster progression time of approximately 3 years. This is over twice as fast as with the nominal diffusion rate, but still within the time span that is reasonable for Parkinsonian progression.

### D. Sensitivity to diffusion rates: reactive oxygen species

For the model of pathogenesis, all oxidative products of Parkinsonian malfunction were amalgamated into one general ROS. To represent the diffusion of ROS we used the value measures for oxygen ions in [Katayama, 2005] of $D_{ros} = 309600$ μm$^2$ h$^{-1}$. In fact, in our model, the diffusion coefficient of ROS has little impact on the progression times. The diffusion coefficient ($D_{ros}$) can be increased or decreased by two orders of magnitude, without any significant difference to the calculated progression times. The prediction is therefore that the progression speed of Parkinson's is largely independent of the diffusion rate of ROS.

## V. DISCUSSION

In this article we have focussed on an example where Parkinsonian damage is initiated in the ENS. This was done for two reasons. First in order to explain how the R-D model of the progression mechanism offers a rational explanation of how invading pathogens can start Parkinson's without continuing with the process – thus resolving the 'Pan-Montojo paradox'. And second to show how the R-D model explains the differing progression rates of Parkinson's disease and its dependence upon the level of risk factors and intrinsic vulnerability.

<u>Other start points for Parkinson's disease</u>

The R-D progression model is of course equally applicable for Parkinson's disease initiated at other start points. For example, external start points such as the olfactory bulb, [Braak 2008] and areas of the nervous systems which are subject to unsupportable energetic loading [Wellstead, 2010, Bolam, 2012] such as the SN [Matsuda 2009, 2012].

## Supplementary Material.

### SM1. The Parkinson's feedback *motif* – a simplified mathematical model of oxidative and protein metabolism.

The pathogenic switching of states displayed by the full mathematical model [Cloutier, 2012a] can be represented accurately in the simplified model – or motif [Cloutier, 2012b] – shown in Figure SMF1. The key feature of the *motif* is a feedback loop, in which the concentration of ROS and αSYN$_{mis}$ are regulated by the endogenous influence of ROS concentration on the production rate ($v_3$) of αSYN$_{mis}$ and a corresponding endogenous influence of αSYN$_{mis}$ concentration on the production rate ($v_1$) of ROS. The influence of external risk factors is coded into the 'input' variables ($S_1$, $S_2$, $S_3$, $S_4$) and how they change the production and removal rates of ROS ($v_1$, $v_2$) and αSYN$_{mis}$ ($v_3$, $v_4$).

The differential equations which describe the *motif*'s dynamical behaviour are given in TABLE SMT1 and the inputs are defined in TABLE SMT2

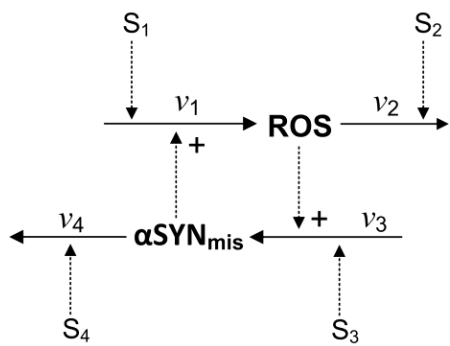

**Figure SMF1**: *Core Feedback Motif for the reaction dynamics of Parkinson's pathogenesis*

**TABLE SMT1** Core feedback motif model

| Symbol | Name | Steady-state value | Differential equations and rates |
|---|---|---|---|
| **State variables** | | **(μM)** | |
| ROS | Reactive oxygen species | 1 | $\dfrac{\partial ROS}{\partial t} = v_1 - v_2$ |
| αSYNmis | Misfolded α-synuclein | 1 | $\dfrac{\partial \alpha SYNmis}{\partial t} = v_3 - v_4$ |
| **Fluxes*** | | **(μM·h$^{-1}$)** | |
| $v_1$ | ROS generation | 468 | $v_1 = k_1 \left[ 1 + S_1 + d_{\alpha SYN} \dfrac{\left(\alpha SYNmis / K_{\alpha SYN}\right)^4}{1 + \left(\alpha SYNmis / K_{\alpha SYN}\right)^4} \right]$ |
| $v_2$ | ROS removal | 468 | $v_2 = k_2 \cdot ROS \cdot S_2$ |
| $v_3$ | αSYNmis generation | 0.007 | $v_3 = k_3 \cdot ROS \cdot S_3$ |
| $v_4$ | αSYNmis removal | 0.007 | $v_4 = k_4 \cdot \alpha SYNmis \cdot S_4$ |
| **Parameters*** * | $k_1 = 468$; $k_2 = 468$; $k_3 = 0.007$; $k_4 = 0.007$; $d_{\alpha SYN} = 15$; $K_{\alpha SYN} = 8.5$; | | |

\* Signals $S_{1-4}$ have the following default values: $S_1 = 0$; $S_{2-4} = 1$.
\** Parameters units are: μM·h$^{-1}$ for $k_1$; h$^{-1}$ for $k_2$, $k_3$, $k_4$; μM for $K_{\alpha SYN}$; dimensionless for $d_{\alpha SYN}$.

| Input | Corresponding pathogenic mechanisms |
|---|---|
| | **TABLE SMT2** Inputs to the core motif and PD pathogenic factors |
| $S_1$ | *Internal and external oxidative stresses*<br>The following mechanisms will act to <u>increase</u> the input $S_1$:<br>I. External oxidative stress from toxins [9, 31]<br>II. Mitochondrial dysfunction [15] including age related decline in efficiency [16, 17] and effects of calcium [5]<br>III. Dopamine induced oxidative stress [5]<br>IV. Elevated free iron level and impaired free radical defence [32] |
| $S_2$ | *Age-related anti-oxidative mechanisms*<br>The following mechanisms will act to <u>decrease</u> the input $S_2$:<br>I. Reduced anti-oxidative capability and GSH buffering with age [33]<br>II. Decline in efficiency of energy metabolism [34, 35]<br>III. Reduced effectiveness of glucose uptake and utilisation (i.e. potentially leading to reduced PPP flux) |
| $S_3$ | *Influence of genetic damage/mutation*<br>The following mechanisms will act to <u>increase</u> the input $S_3$:<br>I. Over expression of the αSYN gene [36]<br>II. Elevated mutation rate of αSYN [36] |
| $S_4$ | *Protein clearance mechanisms*<br>The following mechanisms will act to <u>decrease</u> the input $S_4$:<br>I. *Parkin* gene mutations impairing ubiquitination pathway [36, 22]<br>II. Reduction in available ATP potentially impairs proteins clearance |

## SM2. The reaction-diffusion equations for the progression of Parkinson's

The reaction-diffusion equation for the progression of ROS and αSYN$_{mis}$ along a hypothetical one-dimensional pathway:

$$\begin{bmatrix} \frac{\partial ROS}{\partial t} \\ \frac{\partial \alpha SYN_{mis}}{\partial t} \end{bmatrix} = \begin{bmatrix} v_1 & -v_2 \\ v_3 & -v_4 \end{bmatrix} + \begin{bmatrix} D_{ROS} & 0 \\ 0 & D_{\alpha SYN_{mis}} \end{bmatrix} \begin{bmatrix} \frac{\partial^2 ROS}{\partial x^2} \\ \frac{\partial^2 \alpha SYN_{mis}}{\partial x^2} \end{bmatrix}$$

Where:
  $D_{ros}$ is the diffusion coefficient for ROS
  $D_{αSYNmis}$ is the diffusion coefficient for damaged α-synuclein
  Time is denoted by the variable *t*, and *x* is the position along the pathway. The fluxes $v_1$, $v_2$, $v_3$, $v_4$ are as defined in SM1.

We assumed that an external pathogen has created an initial α-synuclein misfold concentration of 17 μM at start point. This was used as it corresponded to the steady state level of αSYN$_{mi}$ in the model when in the disease state [Cloutier, 2012a]. However, the physiological concentration of α-synuclein varies on the location within a neuron, but is on average 100 μM [Raichur, 2006], and under healthy conditions approximately 1% of the protein is misfolded [Sneppen, 2009]. The concentration of α-synuclein misfolds in Parkinsonian brains has been calculated [Chiba-Falek, 2006] as four times greater than in healthy brains, so that on average the concentration of mis-folded α-synuclein in a Parkinsonian state could be expected to be approximately 4 μM. However, this does not account for variations throughout a neuron and other factors remarked upon by [Chiba-Falek, 2006]. Hence use of an initial α-synuclein misfold concentration of 17 μM is highly conservative.

**Diffusion coefficients:**

ROS

For the reactive species, ROS we used the value of $D_{ros} = 309600$ μm$^2$ h$^{-1}$ as measured for oxygen ions in [Katayama, 2005].

αSYN$_{mis}$

The diffusion coefficient for α-synuclein misfolds is more problematic. Protein diffusion coefficients within a cell have been measured [Kuhn, 2011] as varying between 72,000 μM$^2$/h (in cytoplasm) and 216,000 μM$^2$/h (in cytosol). In the calculation, we used the faster of these two – the cytosolic rate of 216,000 μM$^2$/h  - as the maximum possible speed of diffusion for α-synuclein misfolds. In reality the diffusion rate could be slower than this because: (i) misfolded α-synuclein would diffuse slower than a 'normal' protein, (ii), diffusion would be through a medium that is more 'crowded' that simple cytosol.  Also our simplifying assumption is of one long axonal connection between ENS and brain stem, thus making no allowance for protein transport from one neuron to another [Desplats, 2009].

The physiological concentration of α-synuclein varies on the location within a neuron, but is on average 100 μM [Raichur, 2006], and under healthy conditions approximately 1% of the protein is misfolded [Sneppen, 2009]. The concentration of α-synuclein misfolds in Parkinsonian brains has been calculated [Chiba-Falek, 2006] as four times greater than in healthy brains, so that on average the concentration of mis-folded α-synuclein in a Parkinsonian state could be expected to be approximately 4 μM. However, this does not account for variations throughout a neuron and other factors remarked upon by [Chiba-Falek, 2006]. Therefore, to be conservative in our calculations, we assumed that an external pathogen has created a much greater initial α-synuclein misfold concentration of 17 μM at the trigger point in the ENS.

Calculation of the diffusion times with these parameters indicate that it would take over 300 years for the concentration of protein at the end point in the lower brain stem to exceed 10 μM by diffusion alone. Computing the progression by diffusion over a 65 year period (the approximate age at which risk of Parkinson's begins to become significant) the concentration of protein at the one metre end point is insignificantly small (less than 0.001 μM).

## SM3 Further reading

Brundin P, Li JY, Holton JL, Lindvall O, Revesz T, (2008), Research in motion: The enigma of Parkinson's disease pathology spread. *Nat Rev Neurosci* 9:741–745

Desplats, P, (2009) Inclusion formation and neuronal cell death through neuron-to-neuron transmission of α-synuclein PNAS vol. 106 no. 31, 13010–13015, doi: 10.1073/pnas.0903691106

Lotharius, J. Brundin, P. (2002), Pathogenesis of parkinson's disease: dopamine, vesicles and α-synuclein Nature Reviews Neuroscience 3, 932-942 (December 2002) | doi:10.1038/nrn983

Kühn T, Ihalainen TO, Hyväluoma J, Dross N, Willman SF, et al. (2011) Protein

Diffusion in Mammalian Cell Cytoplasm. PLoS ONE 6(8): e22962. doi:10.1371/journal.pone.0022962

Chiba-Falek, O, Lopex, GJ, Nussbaum, RL, (2006) Levels of alpha-synclein mRNA in sporadic Parkinson disease patients, Movement Disorder, Oct 21 (10): 1703-8, http://www.ncbi.nlm.nih.gov/pubmed/16795004